\documentclass{article}
\usepackage[utf8]{inputenc}
\usepackage{geometry}
\usepackage{amsmath}
\usepackage{xcolor}
\usepackage{caption}
\usepackage{subcaption}
\usepackage{graphicx}
\usepackage{float}
\usepackage{xfrac}
\usepackage[sorting=none,style=nature,backend=bibtex]{biblatex}
\usepackage{geometry}
\DeclareUnicodeCharacter{2009}{{ }}

\usepackage{hyperref} \hypersetup{ colorlinks=true, citecolor=blue, linkcolor=blue, filecolor=magenta, urlcolor=cyan}

\usepackage{booktabs}
\usepackage{multirow}

\usepackage{authblk}

\title{Fluid-Solid Coupled Simulation of Hypervelocity Impact and Plasma Formation}
\author[1] {Shafquat T. Islam}
\author[1] {Wentao Ma}
\author[2] {John G. Michopoulos}
\author[1] {Kevin Wang\thanks{Corresponding author. E-mail address: kevinwgy@vt.edu.}}
\affil[1] {Department of Aerospace and Ocean Engineering, Virginia Tech, Blacksburg, VA, USA}
\affil[2] {US Naval Research Laboratory, Washington DC, USA}

\begin{document}

\maketitle

\begin{abstract}

Previous theoretical and computational studies on hypervelocity impact have mainly focused on the dynamic response of the solid materials that constitute the projectile and the target, while the surrounding environment is often assumed to be a vacuum. In this paper, we consider impact events that occur in a fluid (e.g., gas) medium, and present a computational model that includes the dynamics, thermodynamics, and ionization of the surrounding fluid material. The model couples the compressible inviscid Navier-Stokes equations with the Saha ionization equations. The three material interfaces between the projectile, the target, and the ambient fluid are tracked implicitly by solving two level set equations that share the same velocity field. This method naturally accommodates the large deformation, contact, and separation of interfaces, while avoiding spurious overlapping of different material subdomains. Across the fluid-solid material interfaces, both the state variables (e.g., density) and the thermodynamic equation of state vary significantly. To account for these discontinuities, we compute the mass, momentum, and energy fluxes at material interfaces using the FInite Volume method with Exact multi-material Riemann problems (FIVER). In this method, an exact one-dimensional bimaterial Riemann problem is constructed and solved along each edge in the mesh that penetrates a material interface. The implementation of this computational model is first verified using two example problems for which either the exact solution or reference data are available. Next, the model is utilized to analyze the impact of a tantalum rod projectile onto a soda lime glass (SLG) target in an argon gas environment. In different analyses, the impact velocity is varied between $3$ and $6~\text{km}/\text{s}$, and the radius of the projectile is varied between $2.5$ and $10$ mm. Each analysis starts with a steady-state fluid dynamics simulation that generates the shock-dominated hypersonic flow around the projectile. This flow field is then used as an initial condition to start the fluid-solid coupled impact simulation. The predicted maximum temperature and pressure within the SLG target are found to agree reasonably well with published experimental data for a similar material (fused quartz). Within the ambient gas, a shock wave is generated at the point of collision. It is found to be stronger than the initial bow shock in front of the projectile. The impact simulations also reveal a region of argon gas with high pressure and temperature, formed in the early stage of the impact mainly due to the hypersonic compression of the fluid between the projectile and the target. The temperature within this region is significantly higher than the peak temperature in the solid materials. For impact velocities higher than $4~\text{km}/\text{s}$, ionization is predicted in this region. This finding indicates that the ambient gas may be a contributor to the impact generated plasma for terrestrial and atmospheric applications of hypervelocity impact.
\end{abstract}


\section{Introduction}
Hypervelocity impact is a challenging multiphysics problem that features the rapid transport and dissipation of kinetic energy through mechanical, thermal, chemical, and electromagnetic pathways. In the past, extensive research has been conducted to understand and predict the mechanical response of the target and the projectile, such as shock waves, deformation, fracture, and fragmentation  \cite{schonberg1990hypervelocity, birnbaum_steuben_iliopoulos_michopoulos_2018}. It has also been found that behind the shock waves the solid materials can have pressures of the order of tens of gigapascals, and temperatures of thousands of Kelvins. This extreme thermodynamic state may cause the material to ionize and form a plasma \cite{barsoum, kobayashi1998radiation}. One of the earliest reports of impact-generated plasma was by Friichtenicht and Slattery (1963)~\cite{friichtenicht1963ionization}, in which spherical projectiles made of iron and graphite were accelerated using an electrostatic accelerator to velocities of up to $16~\text{km/s}$. Since then, various authors have investigated the composition and energy of plasma generated under different impact conditions. For example, Ratcliff \textit{et al.} reported plasma temperatures ranging between $20$ eV\footnote{$1~\text{eV} = 11,605~\text{K}.$} and $40$ eV, when the impact velocity is varied between $1.2~\text{km}/\text{s}$ and $87~\text{km}/\text{s}$ \cite{ratcliff_burchell_cole_murphy_alladfadi_1997}. However, the correlation between temperature and impact velocity was not discussed. Also, Lee \textit{et al.} studied impacts on biased and grounded targets. They reported temperature measurements around $2$ eV for impact velocities between $10~\text{km}/\text{s}$ and $30~\text{km}/\text{s}$ \cite{lee_close_goel_lauben_linscott_johnson_strauss_bugiel_mocker_srama,fletcher_close_mathias_2015}. In some cases, it is found that impact generated plasma is accompanied with electromagnetic emissions. One of the earliest reports in this regard was made by Bianchi \textit{et al.} \cite{bianchi_capaccioni_cerroni_coradini_flamini_hurren_martelli_smith_1984}. Since then, some efforts have been made to characterize and explain these emissions. Maki \textit{et al.} conducted rail gun experiments with polycarbonate projectiles with a mass of $1.1$ g, accelerated to $2 - 6.7~\text{km}/\text{s}$~\cite{maki2004radio, maki2005dependence}.  They found electromagnetic emissions in the microwave frequency range (22 GHz-band). They hypothesized that these emissions may come from impact-generated micro-cracks in the target. More recently, Close \textit{et al.} performed impact experiments with electrostatic accelerators, using micro-projectiles (mass between $10^{-16}-10^{-11}\text{g}$) made of iron on to aluminum and tungsten targets. They recorded emissions in the radio frequency range using patch antennas tuned to $315$ MHz and $916$ MHz. The impact velocities for these experiments ranged between $3~\text{km}/\text{s}$ and $66~\text{km}/\text{s}$~\cite{close2013detection}. They attributed the emissions to the coherent motion of charged particles in the plasma due to a self generated ambipolar field \cite{fletcher_close_2017}. Overall, researchers' knowledge on impact-generated ionization, plasma formation, and electromagnetic emission is still limited. Discrepancies amongst different studies can often be attributed to the lack of uniformity in experimental conditions. For example, the pressures in vacuum chambers can be several orders of magnitude higher in light gas gun experiments, which use a gas to accelerate projectiles, compared to experiments done with electrostatic accelerators. While experiments using electrostatic accelerators can attain a better vacuum condition and accelerate projectiles to higher velocities, they can only launch projectiles of a much smaller mass.

Partly due to the high experimental costs and limitations of apparatus, there has been growing interest in developing computational models to predict impact-generated plasma and electromagnetic waves. For example, Li {\it et al.} ~\cite{song_li_ning_2013, li_song_ning_2014} simulated hypervelocity impact of aluminum projectile and target for impact velocities between $5$ and $10~\text{km}/\text{s}$ using both commercial and self-developed codes that couple smooth particle hydrodynamics (SPH) with the Thomas-Fermi model. Fletcher {\it et al.}~\cite{fletcher_close_mathias_2015} also developed a SPH code to simulate hypervelocity impacts and used the non-ideal Saha equations to predict ionization in the target material. Later, Fletcher {\it et al.} developed a  particle-in-cell (PIC) code to investigate the source of electromagnetic emission from the impact generated plasma \cite{fletcher_close_2017}. Recently, La Spina {\it et al.} combined a spherically symmetric blast wave model with the non-ideal Saha equations and the Frank-Tamm formula to investigate the onset of Cherenkov radiation from glass materials under hypervelocity impact \cite{la2023semi}. Despite these progresses, the generation of plasma and electromagnetic waves from hypervelocity impacts remains an active research area. Open questions in this area include the source (i.e.~projectile, target, or the surrounding gas) and composition of plasma, the dependence of plasma energy on impact velocity (cf.~\cite{ratcliff_burchell_cole_murphy_alladfadi_1997,lee_close_goel_lauben_linscott_johnson_strauss_bugiel_mocker_srama}), and the energy and spectrum of the electromagnetic emissions~\cite{fletcher_close_2017}.

Previous theoretical and computational studies on hypervelocity impact usually assume the ambient environment to be a vacuum, thereby neglecting its role in the impact events. This assumption can be valid for impact events that occur on spacecrafts in the absence of an atmosphere. However, the ambient fluid may be a significant contributor to the impact generated plasma for terrestrial and atmospheric applications of hypervelocity impact. Prior to the projectile making contact with the target, it produces a shock-dominated hypersonic fluid flow during its flight. When the projectile impacts on the target, another  shock wave forms at the point of contact, and propagates radially through the fluid medium. This shock wave is much stronger than the bow shock formed during the hypersonic flight of the projectile \cite{islam2023plasma}. Compared to the solid materials that constitute the projectile and the target, the ambient gas has much lower density, and is far more compressible. Therefore, the gaseous material behind the shock wave may reach a temperature that is higher than that in the solids. Hence, the ambient gas may also ionize, thereby contributing to the plasma mixture formed during hypervelocity impact events. It is imperative to understand the role of the ambient fluid material in order to develop a complete description of hypervelocity impacts.

In this paper, we present the development of a fluid-solid coupled computational model of hypervelocity impact, including the formation of plasma within the ambient fluid. The computational domain is comprised of three non-overlapping subdomains, occupied by the projectile, the target, and the ambient fluid, respectively. The inclusion of the ambient fluid flow is a main feature that distinguishes this work from previous studies (e.g.,~\cite{BOUCHEY2020103462, giannaros2019hypervelocity}). The compressible inviscid Navier-Stokes equations are adopted to model the dynamics of all the solid and fluid materials. The fluid flow is dominated by shock waves, and the solid structures exhibit large, complex deformations. Therefore, we solve the governing equations in the Eulerian reference frame, using a high-resolution finite volume method. To track the dynamics of the material interfaces (i.e. subdomain boundaries), we apply the level set method~\cite{zhao2023simulating, osher_fedkiw_2009,sethian_sethian_1999}. In particular, the boundaries of the projectile and the target are represented implicitly as the $0$ level set of two signed distance functions. In this way, we solve two level set equations to track three material interfaces, namely, projectile-target, projectile-fluid, and target-fluid. Across the solid-fluid interfaces, mass density jumps by several orders of magnitude, and the thermodynamic relations (i.e.~equations of state) also differ significantly. This type of discontinuities poses a challenge to the computation of fluxes across material interfaces. Several numerical schemes which are known to perform well for single-phase flows, develop spurious oscillations near material interfaces \cite{johnsen2012preventing}. These oscillations may lead to loss of accuracy and numerical stability. In this work, we compute the mass, momentum, and energy fluxes across material interfaces  using the FInite Volume method based on Exact multi-material Riemann problem (FIVER) \cite{farhat_gerbeau_rallu_2012, wang_lea_farhat_2015}. By constructing and solving an exact bimaterial Riemann problem along each edge in the mesh that crosses a material interface, FIVER explicitly accounts for the change of equation of state across the interface. Previously, FIVER has been validated for several shock-dominated multiphase flow and fluid-structure interaction problems in underwater explosion and implosion, pipeline explosion, cavitation erosion, and shock wave lithotripsy \cite{farhat_wang_main_kyriakides_lee_ravi-chandar_belytschko_2013, wang_lea_farhat_2015,wang_2017,cao2019shock,ma2022computational}. 

To predict ionization in the fluid  subdomain, we solve the ideal Saha equation, coupled with the conservation of charge and the conservation of nuclei~\cite{zaghloul_2004}. These equations are referred to collectively as the Saha equations. Since the mass density of the gaseous material near the point of contact is very low --- an order of magnitude lower than the ambient ($\sim~2\times10^{-4}~\text{g}/\text{cm}^3$) --- the impact-generated plasma has a large Debye length. This minimizes the effects of non-ideality of the argon plasma \cite{fletcher_thesis}. The Saha equations relate the ionization state of plasma with its thermodynamic state, i.e.~pressure and temperature.  This one-way coupling allows us to predict the extent of ionization and compute the distribution of ionization products. This model assumes local thermodynamic equilibrium, which can be justified for predicting the formation and initial expansion of plasma during hypervelocity impacts~\cite{fletcher_close_mathias_2015}.

We present a verification study that includes two simplified model problems relevant to hypervelocity impact. In the first problem, we assume an infinite ideal plasma at constant pressure, with temperature up to $5\times 10^4~\text{K}$. We solve the Saha equations to compute the composition and mean charge of the plasma, and compare the results with reference data provided in Zaghloul {\it et al.}~\cite{zaghloul_bourham_doster_2000}. The second example is a one-dimensional multi-material impact simulation, for which the exact solution can be obtained up to the time that any rarefaction wave reaches a material interface. 

\begin{figure}
\centering\includegraphics[width= 0.85\linewidth]{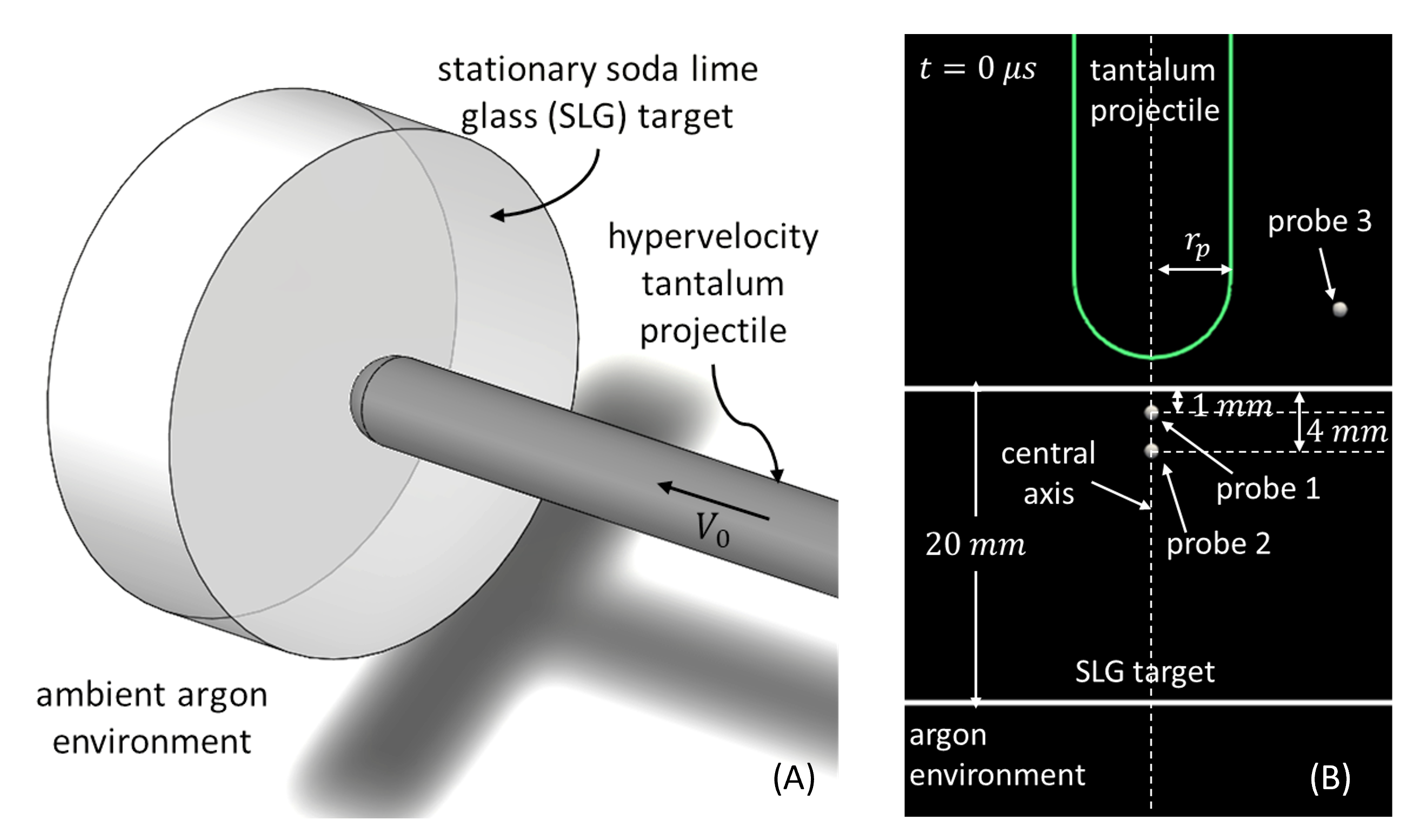}
\caption{ Problem setup. (A) A 3D depiction of the target, projectile, and ambient fluid in the model problem. (B) A 2D cross-section with annotated geometry and probe locations.}
\label{fig: setup}
\end{figure}

Next, we apply the computational model to simulate the impact of a tantalum rod projectile onto a target made of soda lime glass (SLG) in an argon gas environment (Fig.~\ref{fig: setup}). Tantalum is a hard, refractory metal that is often used in impact experiments and applications. SLG is selected as the target material for its potential application in armor and protective systems~\cite{key2020numerical,monroe2021time}. Argon is selected as the ambient fluid  because it is monoatomic, and chemically inert even under extreme pressure and temperature conditions. In different simulations, we vary the projectile's impact velocity ($V_0$) between $3 ~\text{km}/\text{s}$ and $6 ~\text{km}/\text{s}$, and its radius ($r_p$)  between $2.5~\text{mm}$ and $10~\text{mm}$. The velocity and thermodynamic state within the solid and fluid materials are investigated and compared. The extent of ionization in the ambient fluid is characterized by the mean charge and the plasma density.

\section{Materials and Methods} \label{sec: methods}

\subsection{Continuum dynamics}
\label{subsec: continuum dynamics}

Figure~\ref{fig: setup} presents the setup of the hypervelocity impact problem investigated in this paper. The projectile is a cylindrical tantalum rod with a spherical leading edge. The target is a cylindrical plate made of SLG with a radius of $30~\text{mm}$ and a height of $20~\text{mm}$. In the far-field, the density, pressure, and temperature of argon are fixed at $1.78~\text{kg}/\text{m}^3$, $100~\text{kPa}$, and $300~\text{K}$, respectively. Figure~\ref{fig: setup}(B) displays the geometry of the  three material subdomains and the location of three virtual probes that are placed within the SLG target and the ambient gas. 

For the range of impact velocities studied in this work ($3~\text{km}/\text{s}$ to $6~\text{km}/\text{s}$), the density of the energy transferred from the projectile to the SLG target is far greater than the strain energy density of SLG. In the past, Kobayashi {\it et al.}~showed that when SLG is impacted by steel and tungsten projectiles traveling  at  $4~\text{km}/\text{s}$ to $6~\text{km}/\text{s}$, the maximum pressure inside it exceeds $50~\text{GPa}$ \cite{kobayashi1998radiation}, which is an order of magnitude higher than the material's Hugoniot elastic limit ($3~\text{GPa}$ to $8~\text{GPa}$). Tantalum also has a relatively low Hugoniot elastic limit, around $2~\text{GPa}$ \cite{razorenov2012spall}. Therefore, the solid materials involved in this problem, namely tantalum and SLG, are modeled as compressible fluids.

Therefore, the dynamics of the target, the projectile, and the surrounding gas can be considered to be governed by the three-dimensional (3D) compressible Navier-Stokes equations. In addition, we neglect the effects of viscosity and heat diffusion, which reduces the Navier-Stokes equations to

\begin{equation}
\frac{\partial \mathbf{q}}{\partial t}  + \nabla \cdot F(\mathbf{q}) =  0,
\label{eq:Euler}
\end{equation}
In the Cartesian coordinate system,

\begin{gather}
F(\mathbf{q}) = {\begin{bmatrix} f(\mathbf{q}) & g(\mathbf{q}) & h(\mathbf{q})  \end{bmatrix}},\quad\text{with} 
\notag \\
\mathbf{q} = {\begin{bmatrix} \rho \\ \rho u \\ \rho v \\ \rho w \\ E \end{bmatrix} }     ,
\qquad 
f(\mathbf{q}) = {\begin{bmatrix} \rho u \\ \rho u^2+p \\ \rho u v \\ \rho u w \\ \rho H u \end{bmatrix} }   , 
\qquad 
g(\mathbf{q}) = {\begin{bmatrix} \rho v \\ \rho u v \\ \rho v^2+p \\ \rho v w \\ \rho H v \end{bmatrix} }   , 
\qquad 
h(\mathbf{q}) = {\begin{bmatrix} \rho w \\ \rho u w \\ \rho v w \\ \rho v^2 + p \\ \rho H w \end{bmatrix} } .
\notag
\end{gather}

Here, $\rho$ is the mass density. $p$ is the pressure, $\mathbf{V} = {\begin{bmatrix}
u & v & w \end{bmatrix}^T}$ is the velocity vector. $E$ is the total energy per unit volume, given by $\displaystyle E= \rho e + \frac{1}{2} \rho ||\mathbf{V}||^2_2 $; and $H$ is the total enthalpy per unit mass, defined by $\displaystyle H = \frac{1}{\rho}(E+p)$.

Leveraging the cylindrical symmetry of this problem (see Fig.~\ref{fig: setup}), we solve the 3D problem in a 2D computational domain. The 2D governing equations that account for cylindrical symmetry are given by
\begin{equation}
\frac{\partial}{\partial t} {\begin{bmatrix}
    \rho \\ \rho u_r \\ \rho w \\E
\end{bmatrix}}  +  
\frac{\partial}{\partial r} {\begin{bmatrix}
    \rho u_r\\ \rho u_r^2 + p \\ \rho u_r w \\ \rho u_r H
\end{bmatrix}}  +  
\frac{\partial}{\partial z} {\begin{bmatrix}
    \rho w\\ \rho u_r w \\ \rho w^2 + p \\ \rho w H
\end{bmatrix}}
= - \frac{1}{r} {\begin{bmatrix}
    \rho u_r\\ \rho u_r^2 \\ \rho u_r w \\ \rho u_r H
\end{bmatrix}},
\label{eq:Euler cylindrical}
\end{equation}

Here, $u_r$ and $w$ are the radial and axial components of velocity, and $r$ and $z$ are the radial and axial directional coordinates. 

\begin{figure} [h]
\centering\includegraphics[width= 0.85\linewidth]{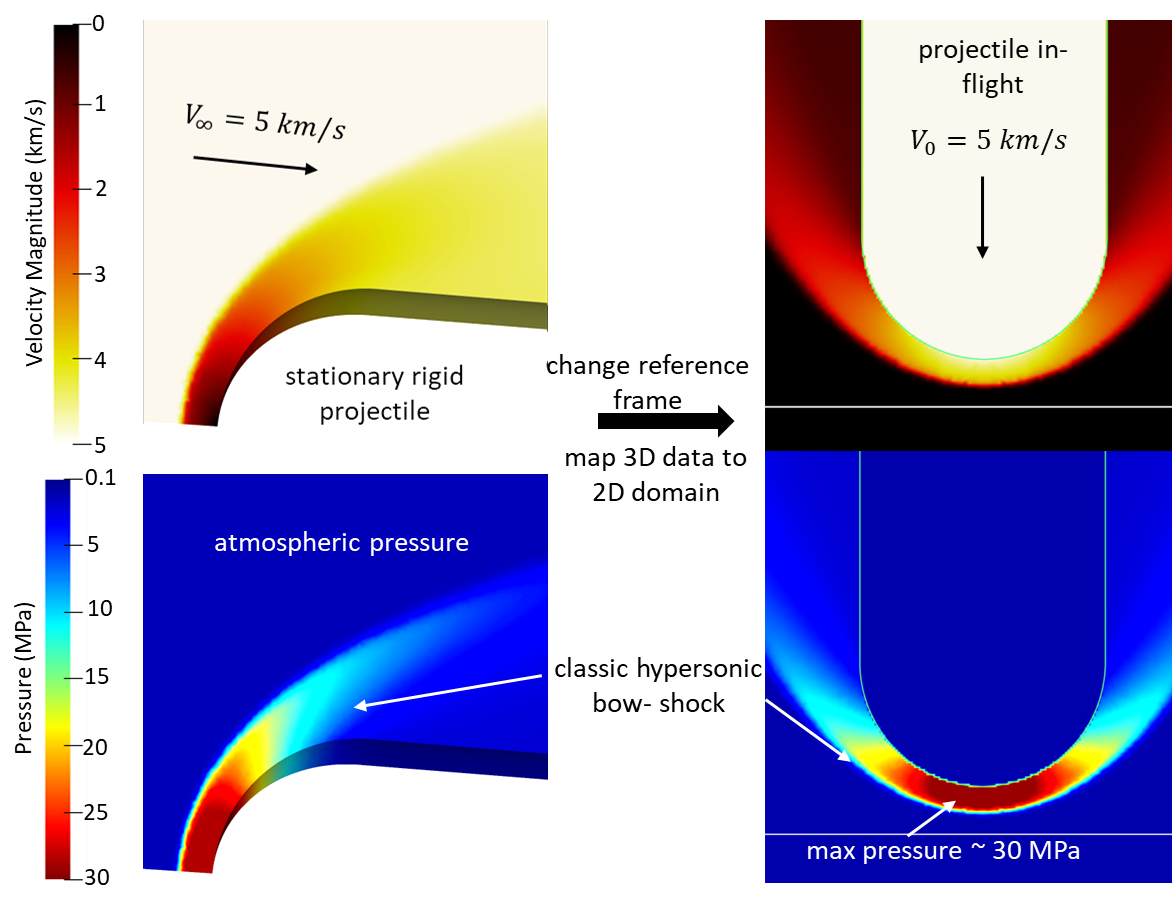}
\caption{ Velocity and pressure in ambient fluid for 3D steady state CFD analysis shown on the left. Mapped data on 2D domain after changing frame of reference --- initial conditions for fluid-solid coupled impact analysis shown on the right}
\label{fig:CFD IC}
\end{figure}

Before the projectile makes contact with the target, the surrounding fluid flow is already non-trivial. It is hypersonic, and features a bow shock that reaches the target before the projectile. To account for this flow field, we perform a 3D steady state computational fluid dynamics (CFD) analysis, and use its result as the initial condition of the fluid-solid coupled impact analysis. This 3D CFD analysis is performed using the AERO-F solver \cite{AERO_F} on an unstructured, body-fitted mesh \cite{huang2020modeling}. In the most refined region, the element size is approximately $0.15~\text{mm}$.  Since this analysis is performed in the inertial reference frame, the far-field fluid velocity $V_\infty$ is given by $V_\infty = -V_0$. As an example, Fig.~\ref{fig:CFD IC}(left) shows the velocity and pressure fields obtained from a CFD analysis with $V_0 = 5~\text{km}/\text{s}$ and $r_p=5~\text{mm}$. The bow shock ahead of the projectile is captured clearly.

After converting to a laboratory reference frame, the pressure, velocity, and density solutions of the  3D CFD analysis are extracted and mapped on to a non-body-fitted 2D Cartesian mesh using radial basis functions. In the most refined region, the element size in the 2D domain is $0.1~\text{mm}$. This mapped data is used as the initial condition for a fluid-solid coupled impact analysis, where the leading edge of the projectile is initialized to be $2 ~\text{mm}$ away from the target surface. This analysis was performed using the M2C solver \cite{M2C}, the results of which is discussed in details in Sec.~\ref{sec: results}. The two solvers used in this study, namely AERO-F~\cite{AERO_F} and M2C~\cite{M2C}, are both publicly available.

\subsection{Material models}

\begin{table*}
\centering
\begin{tabular*}{1\textwidth}{@{\hspace*{1.5em}}@{\extracolsep{\fill}}cccccc@{\hspace*{1.5em}}}

\toprule
Substance & EOS & \multicolumn{4}{c}{Parameters} \\ \midrule
{Tantalum} & {Mie-Gr\"uneisen} & $c_0 \;~(\text{km}/\text{s})$  & $s$ & $\rho_0 \;~(\text{g}/\text{cm}^3)$ & $\Gamma_0$ \\ \cmidrule(l){3-6}
 &  & $3.293$ \cite{mitchell1981shock} & $1.307$ \cite{mitchell1981shock}  & $16.65$ \cite{russell2005structure} & 1.64 \cite{katahara1979pressure}\\ \midrule
{Soda lime glass (SLG)} & {Stiffened gas} & $\gamma$ & $p_c~(\text{GPa})$ &  $c_v~(\text{J}/(\text{K}\cdot\text{kg}))$ &  \\ \cmidrule(lr){3-5}
 &  & $3.9$ & $2.62$ &  $1156$ \cite{huang1992temperature} &  \\ \midrule
{Argon} & {Perfect gas} & $\gamma$ & $c_v~(\text{J}/(\text{K}\cdot\text{kg}))$ &  &  \\ \cmidrule(lr){3-4}
 &  & $1.667$ & $312.2$ &  & \\  \bottomrule
\end{tabular*}
\caption{Parameters of the equations of state}
\label{table:EOS}
\end{table*}

 The compressible Navier-Stokes equations ~\eqref{eq:Euler cylindrical} must be complemented with a thermodynamic equation of state (EOS) to algebraically close the system. In this work, a different EOS is adopted for each material subdomain, to accommodate for their unique features (i.e. silicate target, metallic projectile, and gaseous ambient fluid). 
 
In the literature, the Mie-Gr\"uneisen EOS has been a popular choice to model solid materials in hydrocodes~\cite{zukas2004introduction}. As such, the tantalum projectile subdomain is modeled using this EOS. It can be formulated as~\cite{robinson2019mie}:
\begin{equation}
p(\rho,e) = \dfrac{ \rho_0 c_0^2 \eta}{(1 - s \eta)^2} \Big(1 - \dfrac{1}{2}\Gamma_0\eta\Big) + \rho_0\Gamma_0 e,
\label{eq:MG_EOS}
\end{equation}
where $e$ is the specific internal energy, and $\rho_0$ and $c_0$ denote the density and bulk speed of sound in the ambient condition. $s$ is the slope of the Hugoniot curve. $\Gamma_0$ is the Grüneisen parameter. $\eta$ is the volumetric strain, and can be expressed as: $1 - \rho_0 / \rho$. Argon is modeled using the perfect gas EOS, with specific heat ratio $\gamma = 1.667$. The SLG target is modeled using the stiffened gas EOS~\cite{saurel1999simple}, i.e.
\begin{equation}
p(\rho,e) = (\gamma - 1)\rho e - \gamma p_c,
\label{eq:SG_EOS}
\end{equation}
where $\gamma$ and $p_c$ are empirical model parameters. 

SLG is modeled after the glass commercially known as Starphire\textsuperscript{\textregistered}, which has a chemical composition (by weight): $73\%$ SiO$_2$, $14\%$ Na$_2$O, $10\%$ CaO, and $3\%$ MgO \cite{starphire_vitro}. Unlike tantalum, SLG is not modeled using the Mie-Gr\"uneisen EOS, because certain regions in the SLG target experience high tensile stresses during the impact event. The Mie-Gr\"uneisen EOS can be an excellent choice to model solids under compression, but might lose hyperbolicity when used to model materials under tension. The stiffened gas EOS, however, can be calibrated to capture the shock Hugoniot obtained from laboratory experiments~\cite{saurel1999simple}. Eq.~\eqref{eq:SG_EOS} has been combined with the Rankine-Hugoniot jump conditions, and fit to the shock Hugoniot: $u_s = c_0 + s u_p$. Here $u_s$ and $u_p$ denote the shock speed and the downstream particle velocity, respectively. Using the shock Hugoniot data presented by Grady and Chhabildas ~\cite{grady1996shock} ($c_0 = 2.01~\text{km}/\text{s}$, $s = 1.7$) in the aforementioned procedure, gives $\gamma = 3.9$ and $p_c = 2.62~\text{GPa}$. 

For all the materials, temperature is assumed to be a function of only specific internal energy ($e$). A constant specific heat is specified for each material, which yields a linear relation, $T = (e - e_{0})/c_v + T_0$, where $T$ denotes temperature and the subscript $0$ refers to a reference state. The monoatomic configuration and the absence of valence electrons entail that argon atoms have only translational degrees of freedom, but not vibrational, rotational, or electronic degrees of freedom. Therefore, the specific heat of argon is independent of temperature, which justifies the use of a constant specific heat ~\cite{gyftopoulos_beretta_2005}. For the solid materials, the specific heat is computed using the Dulong-Petit law, which matches reasonably well with measurements obtained in laboratory experiments.  All the material parameters used in the simulations are presented in Table~\ref{table:EOS}.

\subsection{Interface tracking and treatment}

At any time $t\geq 0$, the spatial domain of the fluid-solid 
coupled analysis consists of three material subdomains, occupied by the ambient argon, the tantalum projectile, and the SLG target, respectively. Pressure and normal velocity are continuous across the material interfaces; however, density and the tangential component of velocity may have significant discontinuities. During a hypervelocity impact event, the material subdomains undergo rapid deformation. As such, the motion of the material interfaces are predicted by solving two level set equations that share the same velocity field, $\mathbf{V}$. This method allows us to keep track of three interconnected material interfaces (i.e. projectile-fluid, target-fluid, and projectile-target) that undergo large, complex deformations. In this work, the narrow-band level set method is employed, which means the equations are solved only near material interfaces. 

Specifically, the level set equations are given by
\begin{equation}
\dfrac{\partial \phi_{s}}{\partial t} + \mathbf{V}\cdot\nabla\phi_{s} = 0,~~~~s = 1,2.
\end{equation}

\begin{figure}
\centering\includegraphics[width= 1\linewidth]{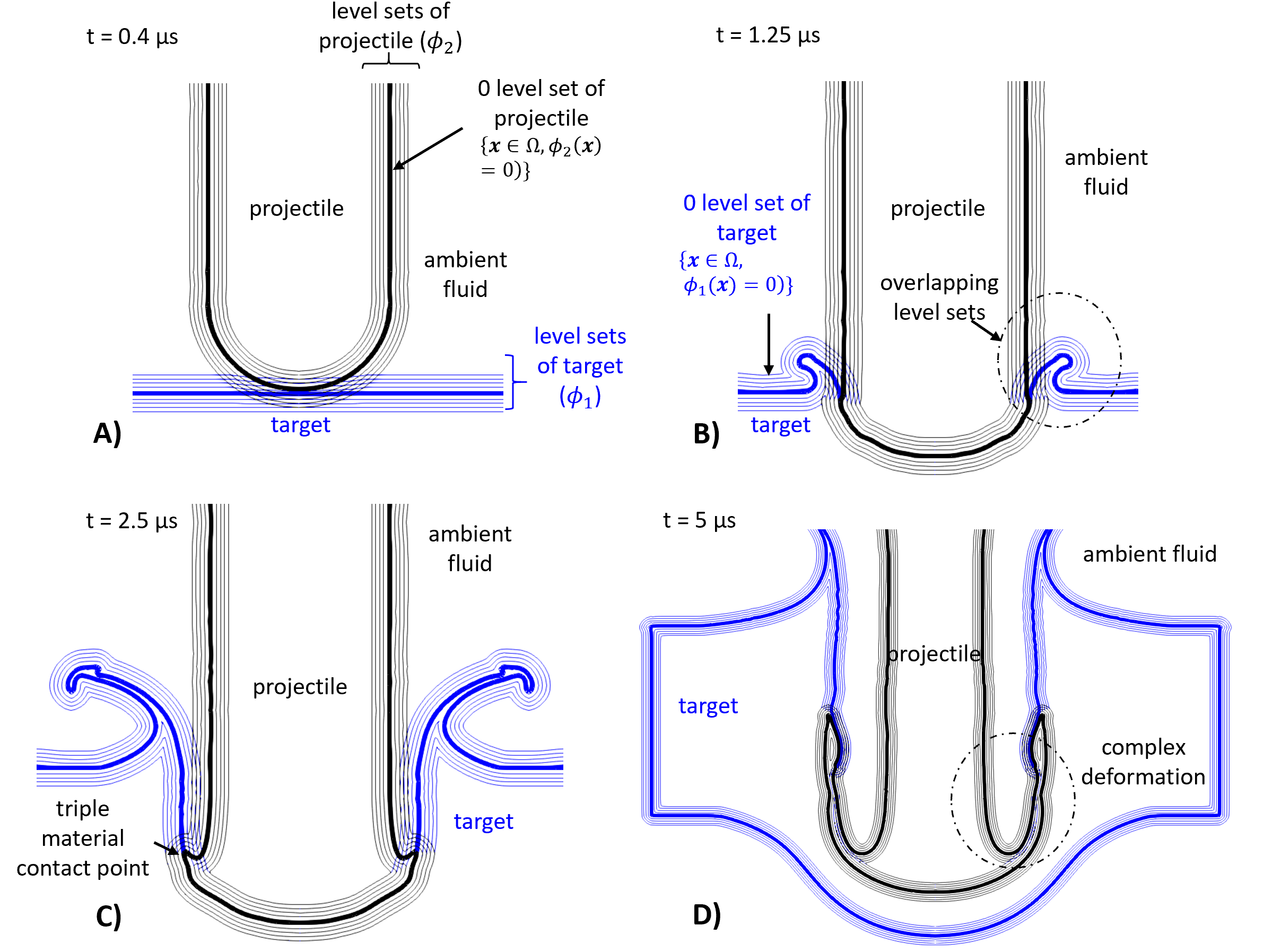}
\caption{Contours of the level set functions for a representative impact simulation ($r_p = 5~\text{mm}$, $V_0 = 5~\text{km}/\text{s}$). A) $t = 0.4~\mu\text{s}$. B) $t = 1.25~\mu\text{s}$. C) $t = 2.5~\mu\text{s}$. D) $t = 5~\mu\text{s}$.}
\label{fig:level set eg}
\end{figure}

Here, $\phi_s$ is the level set function employed to track the boundary of the target  ($s=1$) or the  projectile ($s=2$). $\phi_s$ is initialized to be the signed distance from each point in the computational domain to the material subdomain's boundary. Notably, the two level set equations ($s=1,2$) share the same velocity field, and are solved synchronously to track the motion of material interfaces. This method naturally captures the contact and separation between different materials whilst avoiding non-physical subdomain overlaps. As an example, Fig.~\ref{fig:level set eg} visualizes the two level set functions at four time instances in the impact simulation with $V_0 = 5~\text{km}/\text{s}$ and $r_p = 5~\text{mm}$. Compared to numerical methods that diffuse the interface, the current method is able to capture the sharp interfaces between different solid and fluid materials.

Computing the advective fluxes of mass, momentum, and energy at material interfaces is also challenging, as the EOS varies across the interface. In this work, we adopt the FIVER (FInite Volume method based on Exact multiphase Riemann solvers) method, which is based on the construction and solution of exact bimaterial Riemann problems. Specifically, a one-dimensional bimaterial Riemann problem is constructed along each edge in the mesh that intersects a material interface. This exact Riemann problem is solved iteratively, and its solution is used to compute the local fluxes. For additional details about FIVER, the reader is referred to Ref.~\cite{wang_lea_farhat_2015, main2017enhanced, ma2022computational}.

\subsection{Ionization model}

The plasma density, the molar fraction of each ionic state, and the mean charge number of the plasma are calculated by solving the ideal Saha equation, i.e.
\begin{equation}
\label{eq: Saha}
\frac{n_{r+1} n_e}{n_r} = 2 \frac{U_{r+1}}{U_r} \left[ \frac{2 \pi m_e k_B T}{h^2} \right]^{3/2} \exp\left( \frac{-I_r}{k_B T} \right),~~~~r = 0, 1, ... , N,
\end{equation}
where $n_r$ is the number density of the $r$-th charge state ion, $n_e$ is the number density of electrons (also referred to as the plasma density), $T$ is the temperature, $h$ is the Planck constant, $k_B$ is the Boltzmann constant, $m_e$ is the stationary mass of an electron, and $I_r$ the $r$-th ionization energy. $N = Z-1$, where $Z$ is the atomic number ($18$ for argon).

The Saha equation assumes the condition of local thermodynamic equilibrium. A plasma under this condition must also obey the condition of quasi-neutrality (i.e.~conservation of charge; see \eqref{eq: con electron}), and from conservation of mass also follows that the plasma must observe conservation of nuclei (see \eqref{eq: con nuclei}). Specifically,

\begin{equation}
    \sum_{i = 1 }^{Z} i n_i = n_e,
    \label{eq: con electron}
\end{equation}

\begin{equation}
    \sum_{r= 0 }^{Z} n_r = n_H.
    \label{eq: con nuclei}
\end{equation}

Here, $n_H$ is the number density of ``heavy particles", or nuclei. Following the discussion of Zaghloul \textit{et al.}~\cite{zaghloul_2004}, the combination of Eqs.~(\ref{eq: Saha}, \ref{eq: con electron}, \ref{eq: con nuclei}) yields the one-dimensional transcendental equation,
\begin{equation}
    Z_{av} = \sum_{j = 1}^{J} c_j \times \left[\left(\sum_{i = 1}^{Z_j} \frac{i}{(Z_{av} n_H)^i} \prod_{m = 1}^{i} f_{m, j}\right) \middle/ \left( 1 + \sum_{i = 1}^{Z_j}\frac{\prod_{m = 1}^{i} f_{m, j}}{(Z_{av} n_H)^i} \right) \right]
\label{eq:Zav}
\end{equation}
where $Z_{\text{av}}$ denotes the mean charge in the plasma and $c_j$ the molar fraction of the $j$-th species. $f_m$ is given by
\begin{equation}
f_m = 2\dfrac{U_{m+1}}{U_m} \Big(\dfrac{2\pi m_e k_B T}{h^2}\Big)^{\frac{3}{2}} \exp\Big(-\dfrac{I_m}{k_BT}\Big).
\label{eq: fm}
\end{equation}

 $U_r$ is the state-dependent partition function of the $r$-th charge state ion, given by
\begin{equation}
\label{eq: partition function}
    U_r = \sum_{n = 1}^{n_{\text{max}}} g_{r,n}\exp\left(-\frac{E_{r, n}}{k_B T}\right),
\end{equation}
where $g_{r,n}$ and $E_{r,n}$ denote the degeneracy and excitation energy of the $r$-th ion at the $n$-th energy level. The statistical weighting is done by the degeneracy, $g_{r, n}$ and is related to the angular momentum quantum number, $l_{r, n}$, via the following relation,
\begin{equation}
    g_{r, n} = 2l_{r, n} +1
    \label{eq: l2g}
\end{equation}

The summation in \eqref{eq: partition function} is limited to a maximum excitation state at $n = n_{\text{max}}$, where $n_{\text{max}}$ indicates the last element in the $E_{r,n}$ sequence ($n=1,2,...$) that is smaller than or equal to $I_r$. The values of $I_{r}$, $E_{r,n}$, and $l_{r,n}$ for all elements modelled in this study are obtained from the NIST atomic spectral database~\cite{NIST}. Representative data for the only first three energy levels of the first 4 ions of argon are shown in Table~\ref{table:NIST data}. However, up to $1,685$ energy levels of argon were used in this study.

Eq.~\eqref{eq:Zav} is solved in each time step at each node occupied by argon. This transcendental equation in $Z_{av}$ is solved using a safeguarded iterative method, TOMS748 \cite{alefeld1995algorithm}. After that, contribution of free electrons liberated from the elemental species $j$ to the average charge per heavy particle ($\bar{Z}_{e, j}$), can be computed by evaluating \eqref{eq: Zbar}:
 
 \begin{equation}
    \bar{Z}_{e, j} = c_j \times \left[\left(\sum_{i = 1}^{Z_j} \frac{i}{(Z_{av} n_H)^i} \prod_{m = 1}^{i} f_{m, j}\right) \middle/ \left( 1 + \sum_{i = 1}^{Z_j}\frac{\prod_{m = 1}^{i} f_{m, j}}{(Z_{av} n_H)^i} \right) \right]
    \label{eq: Zbar}
 \end{equation}
 
Moreover, the molar fraction of the neutral atom and each ion state can also be calculated using Eq.~(\ref{eq: alpha0}, \ref{eq: recur}). Let $\alpha_{i, j}$ represent the molar fraction of the $i$-th ion of the $j$-th elemental species. Then,

\begin{equation}
    \alpha_{0, j} = \frac{\bar{Z}_{e_j}}{\sum_{i = 1}^{Z_j} \frac{i}{(Z_{av} n_H)^i} \prod_{m = 1}^{i} f_{m, j}}
    \label{eq: alpha0}
\end{equation}

\begin{equation}
\alpha_{r+1, j}  = \frac{\alpha_{r, j}}{ Z_{av} n_H} f_{r+1},\quad r=1,2,\cdots,Z_j.
\label{eq: recur}
\end{equation}

To accelerate the solution process, we tabulate $U_r,~r=0,1,\cdots,10$ as functions of  $\exp(-1/T)$ at the beginning of the impact analysis. In each time step, we calculate the values of $U_r$ using cubic spline interpolation. A stand-alone solver of the Saha equations can be found at \cite{saha_solver}.


\section{Solver verification} \label{sec: verification}

\subsection{Ideal plasma: The Saha equations solver}
In the initial stage of the impact, the assumption of local thermodynamic equilibrium is justifiable. The hydrodynamics of the impact creates the dominant forces, whereas the effect of ionization on the dynamics of the materials can be neglected. As such, only the density and temperature at each point in the computational domain are required to determine the mean charge, ionic molar fractions, and plasma density. 

We solve the Saha equations for an ideal argon plasma at temperatures between $500$ K and $50,000$ K. $100$ analyses are performed, with a temperature step of $500$ K. The argon pressure is fixed at $100~\text{kPa}$ in all the analyses. The values of material parameters $I_r$, $E_{r,n}$, and $l_{r,n}$ are obtained from the NIST spectral database \cite{NIST}. Some sample parameters are shown in table

\begin{table}[H]
\begin{center}
\begin{tabular}{lllll}
\toprule
$r$                            & $I_r  (\text{eV})$ & $n$ & $l_{r, n}$ & $E_{r, n}(\text{eV})$ \\ \hline
\multicolumn{1}{c}{\multirow{3}{*}{0}} & \multirow{3}{*}{15.759} & 1          &       0    &    0.000        \\ \cline{3-5} 
\multicolumn{1}{c}{}                   &                   & 2          &        2    &     11.548       \\ \cline{3-5} 
\multicolumn{1}{c}{}                   &                   & 3          &       1     &     11.624       \\ \hline
\multirow{3}{*}{1}                     & \multirow{3}{*}{27.630} & 1          &   $\sfrac{3}{2}$         &    0.000        \\ \cline{3-5} 
                                       &                   & 2          &      $\sfrac{1}{2}$      &     0.177      \\ \cline{3-5} 
                                       &                   & 3          &       $\sfrac{7}{2}$      &     16.407       \\ \hline
\multirow{3}{*}{2}                     & \multirow{3}{*}{40.735} & 1          &     2       &     0.000       \\ \cline{3-5} 
                                       &                   & 2          &      1      &     0.138       \\ \cline{3-5} 
                                       &                   & 3          &      0      &     0.195       \\ \hline
\multirow{3}{*}{3}                     & \multirow{3}{*}{59.58} & 1          &      $\sfrac{3}{2}$      &     0.000       \\ \cline{3-5} 
                                       &                   & 2          &      $\sfrac{3}{2}$      &     2.615       \\ \cline{3-5} 
                                       &                   & 3          &       $\sfrac{5}{2}$    &     2.631       \\\bottomrule
\end{tabular}
\end{center}
\caption{Sample spectroscopic data of argon for $r=0,1,2,3$ and $n=1,2,3$ \cite{NIST}.}
\label{table:NIST data}
\end{table}

\begin{figure}[h]
\centering\includegraphics[width= 1\linewidth]{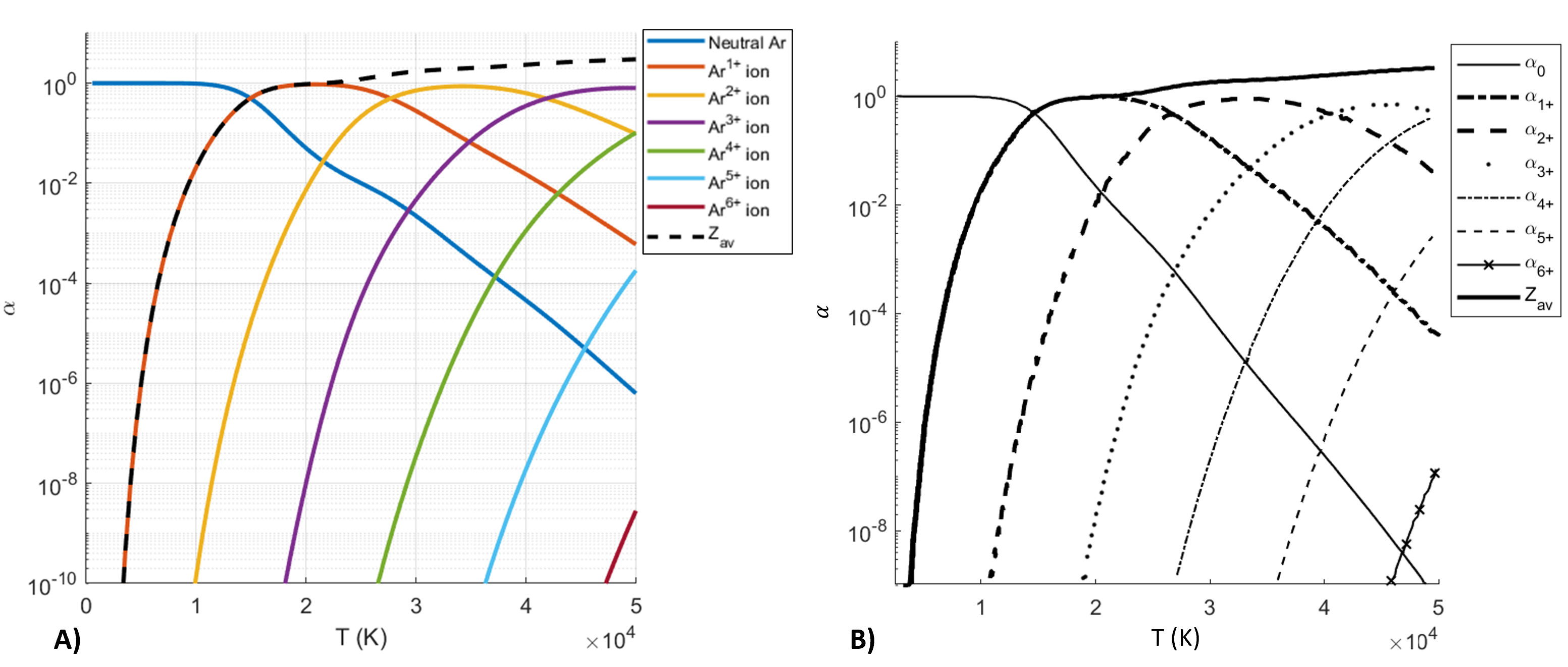}
\caption{Molar fraction ($\alpha$) of ionic species and mean charge ($Z_{\text{av}}$) of ideal argon plasma at 1 atm pressure, for temperature $T \leq 5 \times 10^4$ K. A) Present work. B) Plot recreated from Zaghloul \textit{et al.} \cite{zaghloul_bourham_doster_2000}. }
\label{fig: saha verification}
\end{figure}

Figure~\ref{fig: saha verification} shows the composition of the argon plasma as a function of temperature. It can be observed that as temperature increases, the mean charge number increases monotonically, and the higher ionic species get excited. This solution is in good agreement with results by Zaghloul \textit{et al.} \cite{zaghloul_bourham_doster_2000}. There is some minor discrepancy in the magnitude of the ionic molar fractions in the higher temperature regimes, particularly in the curve for neutral argon. This could be a result of truncation errors or algorithmic accelerations employed in this study. However, the curve for the mean charge shows no such discrepancy and the temperature for onset of each ion is also in excellent agreement with Zaghloul \textit{et al.}

\subsection{1D multi-material hypervelocity impact analysis}
We consider a 1D model of the 3D hypervelocity impact problem illustrated in Fig.~\ref{fig: setup}. The 1D computational domain is along the axial direction of the 3D impact problem ($0 \leq x \leq 1~\text{mm}$). The domain is divided into four separate material subdomains, occupied by argon, tantalum, SLG, and argon again. The initial condition of the 1D model problem is given by
\begin{equation}
\rho(x,0) = 
\begin{cases}
1.78 \times 10^{-3} ~\text{g}/\text{cm}^3 & 0 ~\text{mm} <x \leq 0.15 ~\text{mm},\\
16.65 \text{g}/~\text{cm}^3 & 0.15 ~\text{mm} <x \leq 0.35 ~\text{mm},\\
2.204 \text{g}/~\text{cm}^3 & 0.35 ~\text{mm} <x \leq 0.60 ~\text{mm},\\
1.78 \times 10^{-3} ~\text{g}/\text{cm}^3 & 0.60 ~\text{mm} <x \leq 1 ~\text{mm},\\
\end{cases} 
\end{equation}

\begin{equation}
u(x,0)= 
\begin{cases}
3 ~\text{km}/\text{s} & 0 ~\text{mm} <x \leq 0.35 ~\text{mm},\\
0 ~\text{km}/\text{s} & 0.35 ~\text{mm} <x \leq 1 ~\text{mm},\\
\end{cases}
\end{equation}

and

\begin{equation}
    p(x,0) = 100 ~\text{kPa} \quad 0~\text{mm} \leq x \leq 1~\text{mm}.
\end{equation}

The initial stage of the impact can be represented by a sequence of classical 1D Riemann problems, for which the exact solution can be computed. However, when the impact-generated shock waves reach the back surfaces of the projectile and the target, they reflect as rarefaction fans. These rarefaction fans  propagate towards the projectile-target interface. When either of them reaches the interface, exact solution can no longer be obtained. In this example, the exact solution is available up to $60~\text{ns}$.

\begin{figure*}
\centering\includegraphics[width= 1\linewidth]{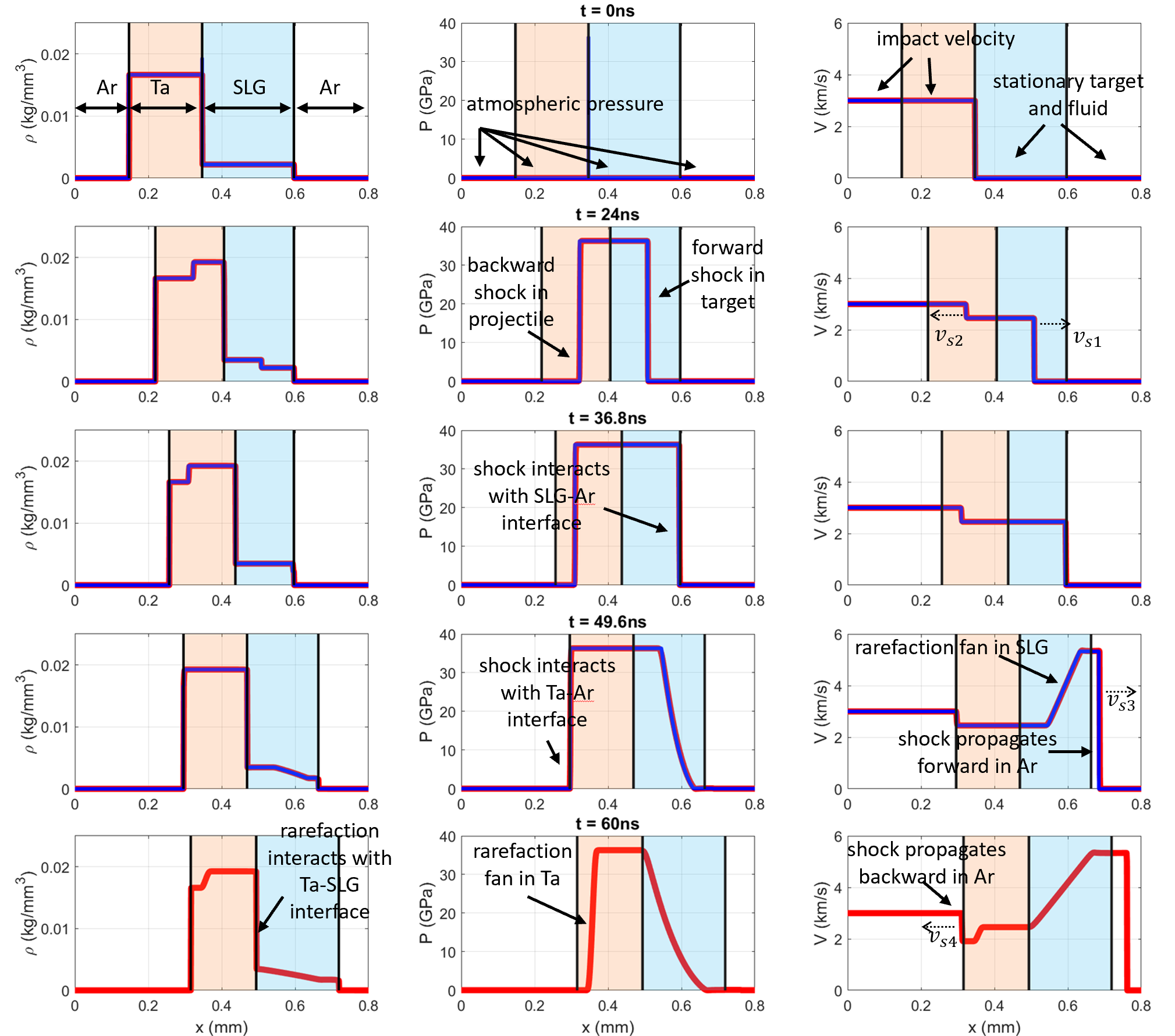}
\caption{Numerical and analytical solution of a one-dimensional hypervelocity impact of a tantalum projectile ($V_0 = 3~\text{km}/\text{s}$) and SLG target in atmospheric argon environment}
\label{fig: 1D HVI}
\end{figure*}

This 1D analysis starts exactly at the time of collision. It is performed up to $60~\text{ns}$. Figure~\ref{fig: 1D HVI} presents the density, velocity, and pressure fields at five time instances: $t = 0 \text{ ns, } 24 \text{ ns, } 36.8 \text{ ns, } 49.6 \text{ ns, and } 60 \text{ ns}$. It can be observed that for all the solution variables, the numerical solution is in excellent agreement with the exact solution.

The initial state is shown in the first row of images in Fig.~\ref{fig: 1D HVI}. The white region denotes the argon domain, the orange color shows the tantalum projectile, and the blue color represents the target SLG. The widths of the solid subdomains were chosen to clearly show the impact dynamics. The projectile and the fluid behind it are initialized at the impact velocity of $3 ~\text{km}/\text{s}$, whereas the stationary target and the fluid behind it is set to be stationary.

The impact sends a forward propagating shock into the projectile and a backward propagating shock into the target, which can be seen in the snapshots taken at $t = 24 \text{ ns}$. The shock waves then propagate through their respective mediums, eventually reaching the solid-fluid material interfaces. The shock wave in the target SLG reaches the SLG-Ar interface first, which sends a rarefaction fan back into the target. The discontinuity in their states at this moment when the shock wave hits the interface can be represented by another Riemann problem, and is shown at $t = 36.8 \text{ ns}$. At this point, the mass density jumps by 3 orders of magnitude across the SLG-Ar interface (i.e. $3.47 ~\text{g}/\text{cm}^3 $ vs. $1.78\times 10^{-3} ~\text{g}/\text{cm}^3 $), which challenges the robustness of the solver. Similarly, when the shock wave in the tantalum projectile reaches Ta-Ar interface, the discontinuity can be represented by yet another Riemann problem, which can be seen at the snap shot at $t = 49.6 \text{ ns}$. This sends a rarefaction fan into the target and a shock into the ambient fluid. However, after $t = 60 \text{ ns}$ the rarefaction fan in the target material reaches the Ta-SLG interface, and beyond this point an analytical solution cannot be obtained, but the system can be numerically solved to simulate the impact further in time.

\begin{figure*}
\centering\includegraphics[width= 1\linewidth]{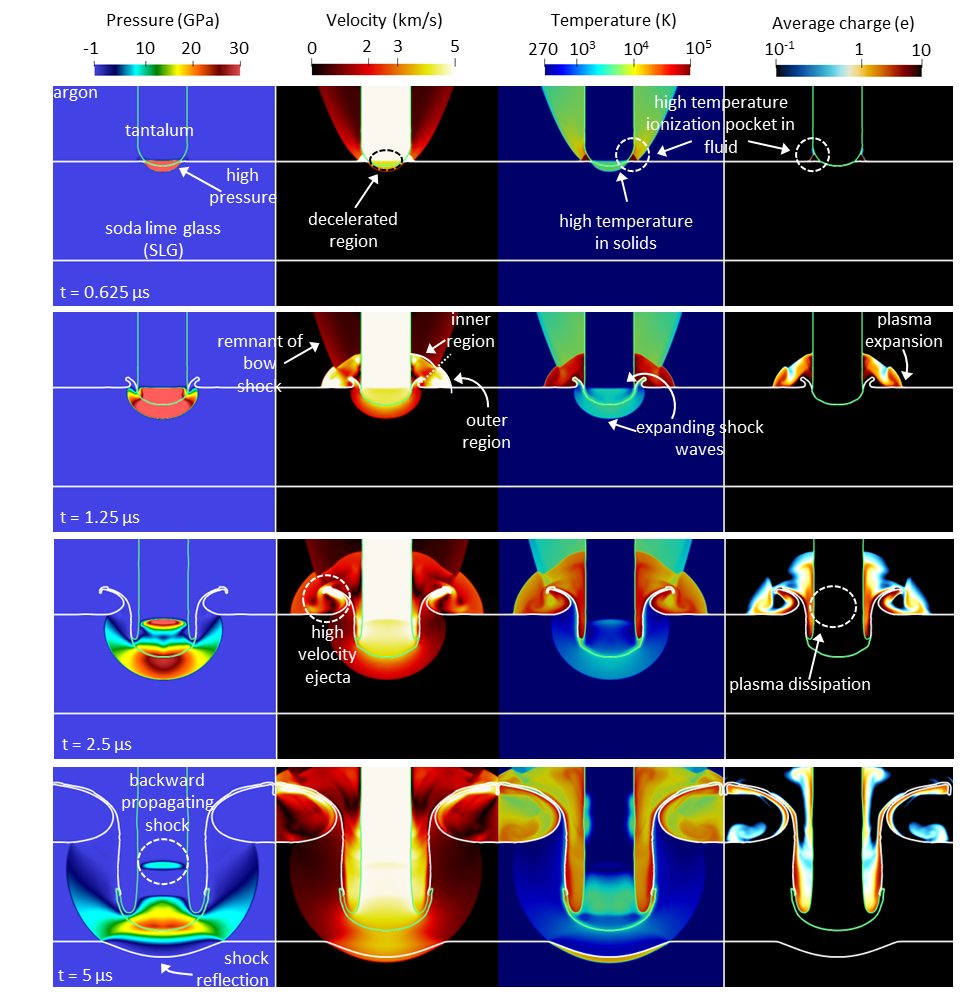}
\caption{Solution snapshots obtained from a representative impact simulation ($V_0 = 5~\text{km}/\text{s}$).}
\label{fig: impact sequence}
\end{figure*}

\section{Results and Discussion} \label{sec: results}

The test case with impact velocity $V_0 = 5~\text{km}/\text{s}$ and projectile radius $r_p = 5~\text{mm}$ can be considered as a representative case. At time $t = 0~\mu\text{s}$, the leading edge of the tanalum projectile is $2$ mm away from the SLG target. Impact occurs at $t = 0.4~\mu\text{s}$. Figure~\ref{fig: impact sequence} presents a sequence of snapshots of the simulation results, taken at $t = 0.625~\mu\text{s, }1.25~\mu\text{s, }2.5~\mu\text{s, and }5~\mu\text{s}$, respectively. The columns in this figure (from left to right) represent the pressure, velocity magnitude, temperature, and mean charge fields. Since the temperature and mean charge values span several orders of magnitude, they have both been plotted on a logarithmic scale. 

\begin{figure}
\centering\includegraphics[width= 0.75\linewidth]{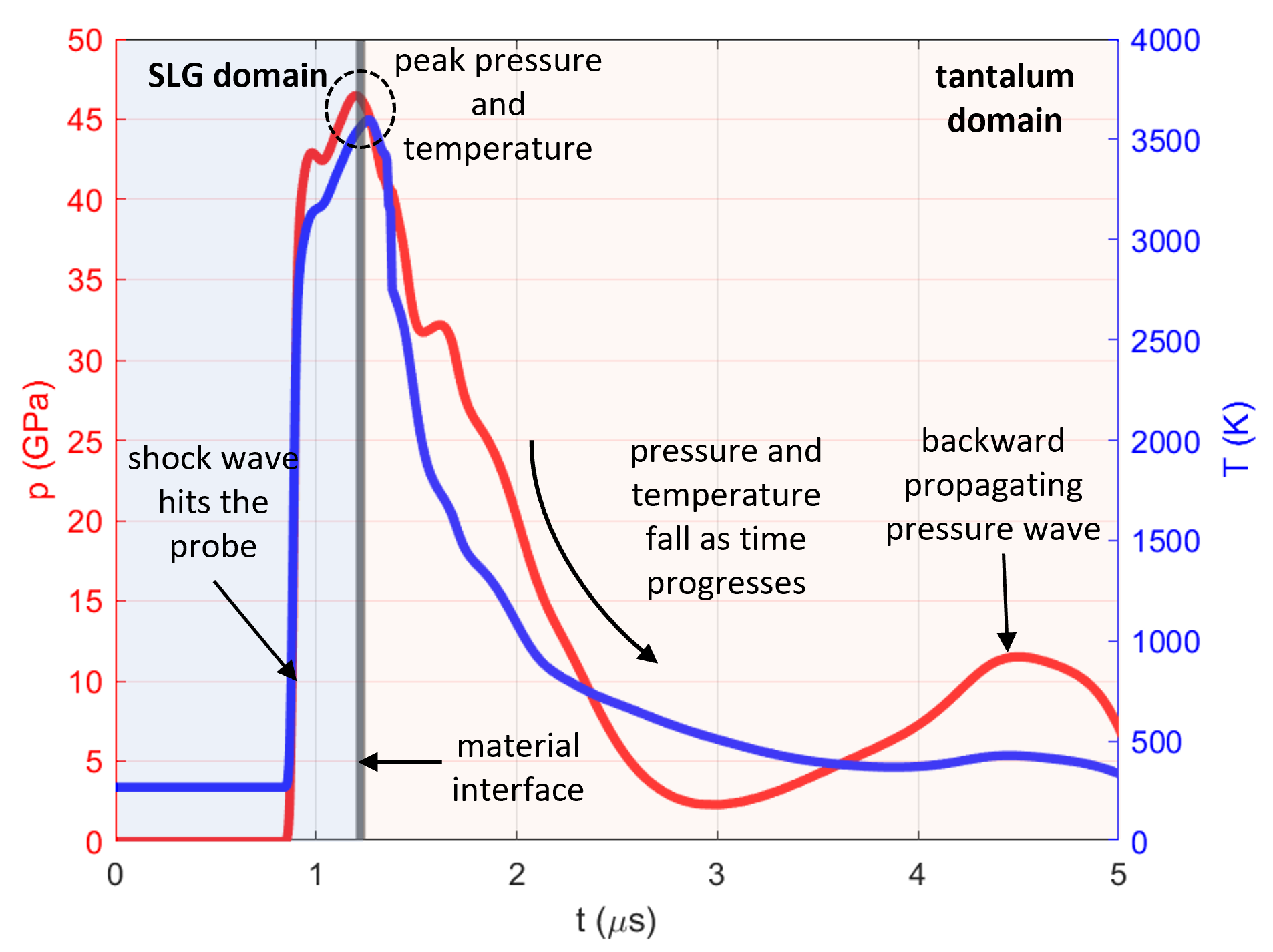}
\caption{Time history of pressure and temperature at probe 2 for $V_0 = 5~\text{km}/\text{s}$ }
\label{fig: probe time history}
\end{figure}

The sudden deceleration of the tantalum projectile due to the collision imparts a large amount of energy from the projectile into the target and the surrounding fluid. This decelerated region can be seen in the velocity field plotted at $t = 0.625~\mu\text{s}$. The impact also causes the formation of high pressure and temperature shock waves within all the three material subdomains. The shock wave propagates radially forward in the target, backwards in the projectile, and outwards from the point of impact in the fluid. However, due to the difference in magnitudes of pressure between the solid and fluid subdomains, only the variations in the projectile and target is visible in Fig (\ref{fig: impact sequence}). The high temperature and pressure in the fluid causes the argon atoms to ionize, and form a pocket of plasma.

At $t = 0.625~\mu\text{s}$ the projectile has displaced a small volume of the target material, and a crater has begun to form. The high pressure region behind the rim of the crater causes the target material to accelerate outwards at approximately $5~\text{km}/\text{s}$ ($V_0$). The pocket of plasma has expanded, and is divided into an inner and outer region by the tip of the SLG ejecta. The ejecta accelerates the fluid in the outer region and it reaches velocities of $ {\displaystyle \sim } 7~\text{km}/\text{s}$, whereas the fluid in the inner region has velocity magnitudes of $ {\displaystyle \sim } 1~\text{km}/\text{s}$. The remnants of the bow shock can still be seen at this time instance. However, the magnitudes of the state variables behind the bow shock are much smaller in comparison to the magnitudes behind the impact generated shock waves; as such the effects of the bow shock on dynamics of the impact are negligible. The temperature behind the shock wave is high in both the solid materials and the argon gas. However, because the argon gas has low density and specific heat, its temperature is several orders of magnitude higher that those found in the solid materials.

As time progresses, the shock waves expand further and the energy density behind them decreases. This process can be clearly seen in the snapshots at $t =2.5~\mu\text{s}$ and $5~\mu\text{s}$. The ionized plasma that was pushed by the ejecta into the outer region has dissipated significantly faster than in the inner region, as the mean charge is much higher within the crater. The pressure wave expands within the target and eventually hits the back wall and causes it to deform radially outwards, and then reflects backwards. The reflected wave destructively interferes with the incident wave, and causes the magnitude of the pressure to fall.

\begin{figure}
\centering\includegraphics[width= 0.75\linewidth]{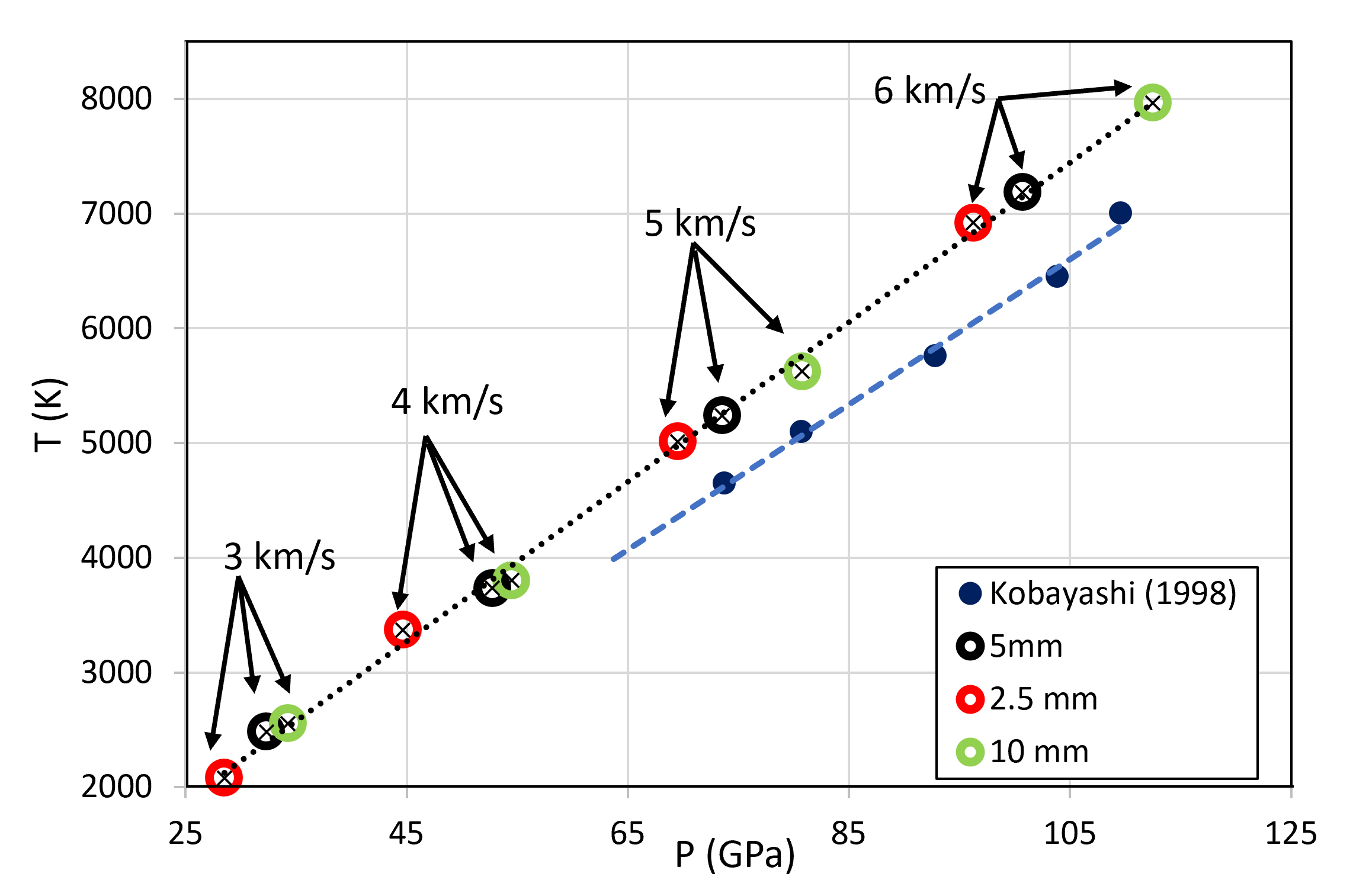}
\caption{Shock temperature vs. shock pressure values at Probe 1, in comparison with experimental results presented in Kobayashi \textit{et al.} \cite{kobayashi1998radiation} }
\label{fig: kobayashi comp}
\end{figure}

For this representative case ($V_0 = 5~\text{km}/\text{s}$, $r_p = 5~\text{mm}$), the time history of pressure and temperature at the Probe 2 are shown in Fig.~\ref{fig: probe time history}. The locations of all the three probes can be found in Fig.~\ref{fig: setup}(B). They are fixed in space during the impact process. Probe 2 is initially within the SLG target, $1$ mm from the upper surface. It is crossed by the leading edge of the tantalum projectile at $1.225~\mu\text{s}$. The time intervals in which the probe is located within SLG and tantalum are shaded in light blue and light red colors in Fig.~\ref{fig: probe time history}. It can be observed that as the impact-generated shock wave reaches the probe location, both  pressure and temperature increase drastically. They keep increasing until the leading edge of the tantalum projectile reaches the probe. Afterwards, both quantities decrease, as the initial forward propagating shock wave propagates and dissipates. A secondary peak is seen in the pressure time history at $t = 4.5~\mu\text{s}$, whereas the temperature does not rise significantly at that time. This occurs as the backward propagating shock wave, which was formed in the projectile at the instance of impact, reaches the probe. The peak pressure and temperature at this probe location are found to be approximately $46~\text{GPa}$ and $3500~\text{K}$, respectively.

Additional simulations are performed with impact velocity $V_0 = 3~\text{km}/\text{s}$, $4~\text{km}/\text{s}$, $5~\text{km}/\text{s}$, and $6~\text{km}/\text{s}$ and $r_p = 2.5~\text{mm}$, $5~\text{mm}$, and $10~\text{mm}$. In each simulation, the maximum pressure and temperature at Probe 1 are extracted. As expected, these peak values are achieved at the initial impact point. It is found that for each impact velocity tested, the projectiles with larger radii generate stronger shocks, that is, with higher magnitude of pressure and temperature behind the shock wave. This behavior is expected, since a greater amount of kinetic energy is deposited from the projectile into the target SLG. Across all the simulations, our computational model predicts a linear dependence of pressure on temperature,  as a linear regression of the $P-T$ values yield an $R^2$ of $0.998$. These values are plotted in Fig.~\ref{fig: kobayashi comp}, alongside the experimental results for one-dimensional impact on fused quartz presented by Kobayashi \textit{et al.} \cite{kobayashi1998radiation}, for a similar range of impact velocities. Probe 1 was placed near the impact surface in the SLG to minimize the three-dimensional effects and to more accurately capture the initial shock state in the SLG. The computational and experimental results agree reasonably well with each other. The slopes of the linear fit of the data differ by less than $9\%$. There is some discrepancy in the intercepts of the linear fits, which may be attributed to the modeling error due to the different compositions of fused quartz and Starphire SLG.

\begin{figure}
\centering\includegraphics[width= 0.75\linewidth]{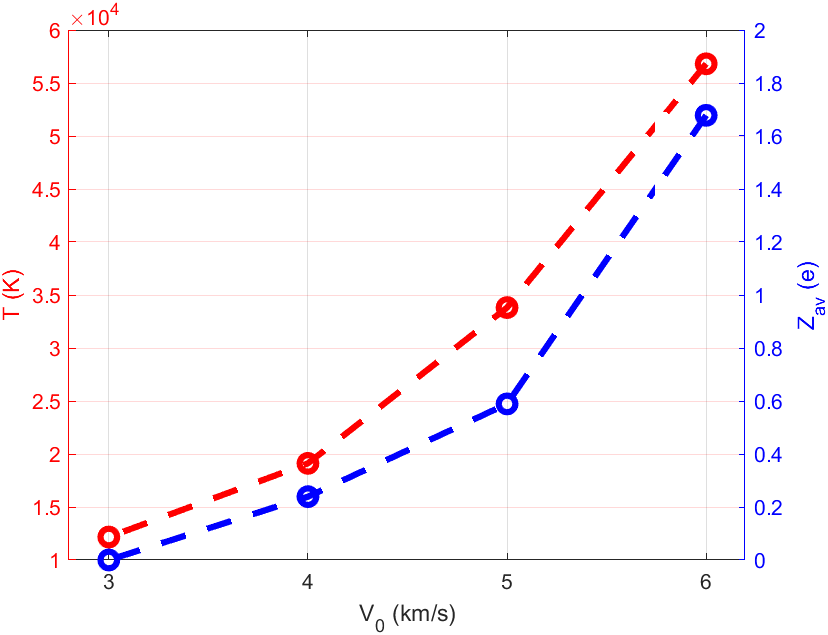}
\caption{Maximum values of temperature ($T$) and mean charge number ($Z_{\text{av}}$) within the ambient argon gas for $V_0 = 5~\text{km}/\text{s}$.}
\label{fig: fluid probe}
\end{figure}

 Figure~\ref{fig: fluid probe} shows the maximum values of temperature and mean charge number obtained at Probe 3, which is placed within the ambient argon gas. As expected, as impact velocity increases, both quantities increase accordingly. Comparing this figure with Figs.~\ref{fig: probe time history} and~\ref{fig: kobayashi comp} shows that the temperature in the ambient gas is significantly higher than that in the solid materials. This is not surprising as temperature is a measure of internal energy per unit mass, and the mass density of the argon gas is much lower than that of the solid materials. Therefore, although the fraction of impact energy (i.e.~kinetic energy carried by the projectile)  transferred to the surrounding gas is small compared to that shared between the solid projectile and target, it is enough to cause dramatic temperature increase in the gas. The high temperature causes argon gas to ionize. Figure~\ref{fig: fluid probe} shows that at the probe location, ionization is significant for impact velocities higher than $4~\text{km}\text{s}$.

\begin{figure}
\centering\includegraphics[width= 0.75\linewidth]{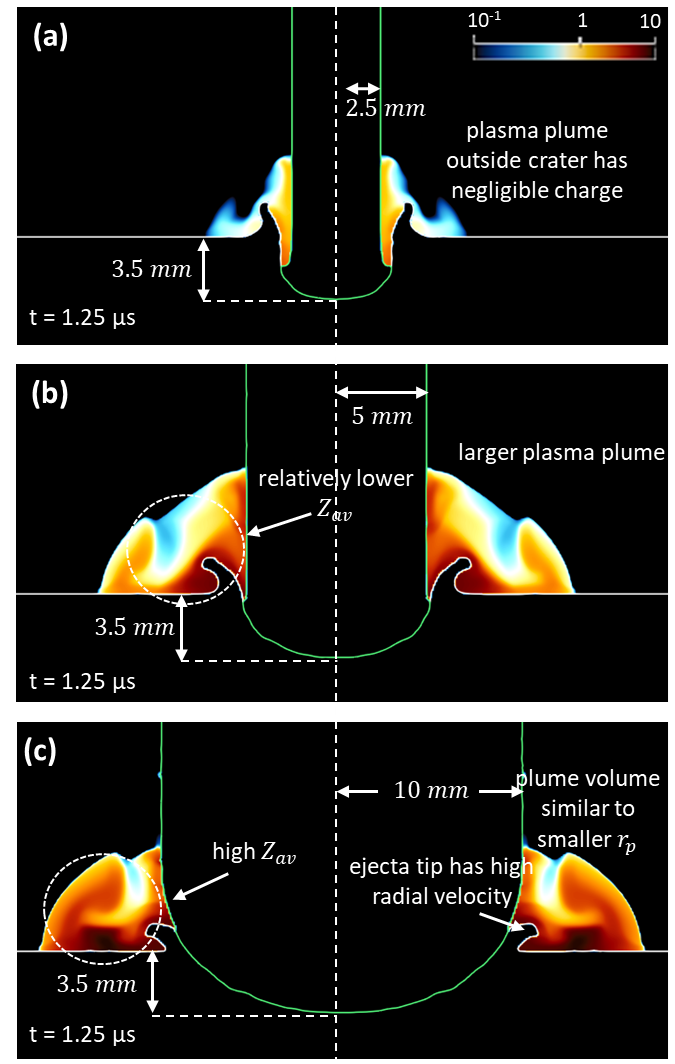}
\caption{Mean charge of impact generated plasma for variable projectile radii.}
\label{fig: Zav for variable rp}
\end{figure}

 Moreover, Fig.~\ref{fig: Zav for variable rp} shows the mean charge, $Z_{\text{av}}$, and the structure of the plasma plume  at $t = 1.25~\mu\text{s}$, obtained from three projectiles with (from top to bottom) $r_p = 2.5~\text{mm, } 5~\text{mm, and } 10~\text{mm}$, while the impact velocity is fixed at $V_0 = 5~\text{km}/\text{s}$. The three projectiles generate similar penetration depth of $3.5 ~\text{mm}$, at this time instance. However, the smaller the projectile radius the more deformation that can be seen in the projectile itself. The structure of the ejecta also displays significant difference, as the projectile with the larger radius has a greater radial velocity at the ejecta tip, but there is limited variance in the axial velocity. This causes the $2.5 ~\text{mm}$ projectile to be expelled more in the axial direction and disrupt the shock structure in the fluid. For the larger radii, the shock structure is still undisturbed at this time, and the shock velocity does not depend significantly on the radius. Therefore, the volume of fluid behind the shock, and thus the volume of the plasma plume, are similar for these cases. However, since the $10 ~\text{mm}$ projectile deposits kinetic energy into the fluid at a faster rate, the extent of ionization (mean charge) is greater than in the $5 ~\text{mm}$ projectile.
 
\section{Conclusion}\label{sec: conclusion}

This paper presents a new computational model of hypervelocity impact that accounts for the dynamics, thermodynamics, and ionization of the ambient fluid (gas), as well as the interaction of the fluid flow with the solid projectile and target. The main features of this model include (1) the solution of two level set equations to track the three sharp material interfaces between the projectile, the target, and the ambient fluid, (2) the construction and solution of exact, one-dimensional bimaterial Riemann problems to enforce interface conditions (a method known as ``FIVER''), and (3) the solution of Saha equations to predict ionization and plasma density within the ambient fluid. The implementation of this model is first verified using two benchmark problems for which either the exact solution or reference data are available. Next, the computational model is applied to simulate the impact of tantalum projectiles on soda lime glass (SLG) in an argon gas environment. In different simulations, the impact velocity is varied between $3$ and $6~\text{km}/\text{s}$, while the radius of the projectile is varied between $2.5$ and $10$ mm. The predicted maximum temperature and pressure within SLG agree reasonably well with published experimental data for a similar material (fused quartz). The temperature in the surrounding gas is found to be significantly higher (by $1\sim 2$ orders of magnitude) than that in the solid materials. This indicates that for impact events that occur in a fluid environment, the fluid may have a substantial effect on the generation of plasma and the emission of electromagnetic waves. For the test case simulated in this paper, ionization of argon is observed at impact velocities above $4~\text{km}/\text{s}$.

\section*{Acknowledgment}
S.T.I, W.M., and K.W.  gratefully acknowledge the support of the Office of Naval Research (ONR) under award N00014-19-1-2102. K.W. also acknowledges the support of the National Science Foundation (NSF) under award CBET-1751487. J.G.M. gratefully acknowledges the support of the ONR under award N00014-21-WX01554.

\printbibliography
\end{document}